\documentclass[12pt, twocolumn]{iopart}
\usepackage[ansinew]{inputenc}
\usepackage{amsfonts}
\usepackage{amssymb}
\usepackage{graphicx}
\usepackage{subeqn}
\usepackage{hyperref}
\usepackage{mciteplus}
\begin{document}
\newcommand{\text}[1]{{\mathrm{#1}}}
\newcommand{\ssl}{\textsc{ssl}}
\newcommand{\ssls}{\textsc{ssl}s}
\newcommand{\sn}{\textsc{sn}}
\newcommand{\dc}{\textsc{dc}}
\newcommand{\ac}{\textsc{ac}}
\newcommand{\iv}{\textsc{iv}}
\newcommand{\ndc}{\textsc{ndc}}
\newcommand{\anc}{\textsc{anc}}
\newcommand{\ie}{{\emph{i.e.}}}
\newcommand{\eg}{{\emph{e.g.}}}
\newcommand{\eqref}[1]{(\ref{#1})}
\newcommand{\eqsref}[1]{(\ref{#1})}
\newcommand{\kb}{k_\text{b}}
\newcommand{\vet}{v_\text{ET}}
\newcommand{\Ibias}{I_\text{dc}}
\newcommand{\iopt}{i_\text{opt}}
\newcommand{\symm}{\mathcal{S}}
\newcommand{\symmto}{\stackrel{\symm}{\to}}
\newcommand{\phiext}{\phi_\text{ext}}
\newcommand{\dphiext}{{\dot \phi}_\text{ext}}
\newcommand{\eps}{\varepsilon}
\newcommand{\uext}{u_\text{ext}}
\newcommand{\weq}{w_\text{eq}}
\newcommand{\omegapl}{\omega_\text{pl}}
\newcommand{\omegas}{\omega_\text{s}}
\newcommand{\gammav}{\gamma_\text{v}}
\newcommand{\gammaw}{\gamma_\text{w}}
\newcommand{\tauv}{\tau_\text{v}}
\newcommand{\tauw}{\tau_\text{w}}
\newcommand{\udc}{u_\text{dc}}
\newcommand{\vdc}{v_\text{dc}}
\newcommand{\vd}{v_\text{r}}
\newcommand{\Eext}{{E_\text{ext}}}
\newcommand{\Weq}{{W_\text{eq}}}
\newcommand{\Ca}{C_\text{a}}
\newcommand{\Cs}{C_\text{s}}
\newcommand{\Rc}{R_\text{c}}
\newcommand{\unit}[1]{{\mbox{#1}}}    
\newcommand{\D}{\mathrm{d}}     
\newcommand{\E}{\mathrm{e}}     
\newcommand{\I}{\mathrm{i}}     
\newcommand{\pd}[2]{\frac{\partial #1}{\partial #2}}    
%
%
%
\title[{THz}-rectification in semiconductor superlattices in the absence of domains]
      {Rectification of {THz}-radiation in semiconductor superlattices
        in the absence of domains}
\date{\today} 
\author{J Isoh\"{a}t\"{a}l\"{a}$^1$ and K N Alekseev$^{1,2}$} 
\address{$^1$ Department of Physical Sciences, P.O. Box 3000, University of Oulu FI-90014, Finland}
\ead{jukka.isohatala@oulu.fi}
\address{$^2$ Department of Physics, Loughborough University LE11 3TU, United Kingdom} 
\begin{abstract}
  We study theoretically the dynamical rectification of a terahertz
  \ac\ electric field, \ie\ the \dc\ current and voltage response to
  the incident radiation, in strongly coupled semiconductor
  superlattices. We address the problem of stability against electric
  field domains: A spontaneous \dc\ voltage is known to appear exactly
  for parameters for which a spatially homogeneous electron
  distribution is unstable. We show that by applying a weak direct
  current bias the rectifier can be switched from a zero \dc\ voltage
  state to one with a finite voltage in full absence of domains. The
  switching occurs near the conditions of dynamical symmetry breaking
  of an unbiased semiconductor superlattice. Therefore our scheme
  allows for the generation of \dc\ voltages that would otherwise be
  unreachable due to domain instabilities. Furthermore, for realistic,
  highly doped wide miniband superlattices at room temperature the
  generated \dc\ field can be nearly quantized, that is, be
  approximately proportional to an integer multiple of $\hbar \omega /
  e a$ where $a$ is the superlattice period and $\omega$ is the
  \ac\ field frequency.
\end{abstract}
\pacs{05.45.-a, 73.21.Cd, 73.40.Ei}
\submitto{\JPCM}
\maketitle
%
%
\section{Introduction}
Semiconductor superlattices (\ssls) have been the subject of much
research ever since Esaki and Tsu\cite{esakitsu70} realized that these
nanostructures could exhibit Bloch-oscillations\cite{bloch28} at
moderate electric field strengths. This oscillatory response makes
\ssls\ inherently nonlinear, and the variety of transport phenomena
that has been discovered has lead to the point that they can now be
called model systems for nonlinear transport\cite{wacker02:ssreview}.
Semiconductor superlattices hold great promise as amplifiers and
detectors of THz radiation. A considerable amount of research has been
put into the problem of THz gain with promising findings both in
theory\cite{hyart08, hyart09_quasistatic, hyart09_magnetic} and in
experiments\cite{savvidis04, sekine05, unuma10}.  Quantum direct
detection\cite{tuckerrev, ignatov99, ignatov02} by THz photon induced
reduction in current has been experimentally
observed\cite{winnerl97:hf-minib, schomburg00, klappenberger01} and
then applied to ultrafast detection and autocorrelation of short THz
pulses\cite{winnerl98, *winnerl99}. A further topic on \ssls\ is
strongly nonlinear dynamical phenomena that arise if one considers for
example circuit models of \ssl\ devices that incorporate effects that
come from the interaction of the charge carriers.  This gives rise to
novel effects such as dissipative
chaos\cite{alekseev96:dissp-chaos-ssl, alekseev02a}, and spontaneous
generation of quantized \cite{dunlap93:bo-f2v-conv,ignatov95,
  alekseev98:spont-dc, romanov00, romanov01}, fractionally quantized,
and non-quantized\cite{romanov00, cannon01-plain} \dc\ bias. This
spontaneous generation of a \dc\ field is also the topic of this
paper.

Spontaneous generation of \dc\ bias is the dynamical effect whereby a
direct electric field appears due to a pure \ac\ excitation. A
significant point is that this spontaneous \dc\ bias can in \ssls\ be
nearly quantized, meaning that the total \dc\ electric field,
$E_\text{dc}$ follows the relationship
\begin{equation}
  E_\text{dc} = \eta n \frac{\hbar \omega}{a e},
  \label{eq:quant}
\end{equation}
where $n$ is an integer, $\omega$ is the frequency of the incident
radiation field and the coefficient $\eta$ represents the fractional
deviation from perfect quantization, $\eta = 1$. The remaining
constants are $a$, $\hbar$, and $e$, which are the superlattice
period, the reduced Planck constant, and the electron charge,
respectively.  This phenomena bears close resemblance to the inverse
\ac-Josephson effect\cite{langenberg, levinsen}, and is in fact just
one example of similarity between Josephson junctions and
\ssls\cite{ignatov93, alekseev02a, isohatala10}. Unquantized
spontaneous \dc\ has also been predicted for lateral semiconductor
superlattices\cite{alekseev05:lsslcr, isohatala10} that only differ by
their geometry from the bulk \ssls\ studied here. Similar effect has
been earlier described in some models of homogeneous bulk
semiconductors\cite{bumyalene89, *bumyalene89-trans}.

We emphasize that this type of rectification occurs in perfectly
symmetric structures, unlike classical rectification that relies on
transport asymmetry due to contact charge inhomogeneities or
nonlinearity. Rather, here a static \dc\ field is formed dynamically
due to presence of absolute negative conductivity (\anc) at zero
\dc\ bias\cite{dunlap93:bo-f2v-conv, ignatov95}. \anc\ in itself, is a
well-known effect in \ssl s\cite{pavlovich76-orig, *pavlovich76nt,
  ignatov76, keay95}, and has been theoretically found to be a robust
effect in \ssl s, occurring for a variety of miniband
dispersions\cite{romanov09} and external excitations\cite{romanov04,
  romanov06, alekseev06}.

In \ssls\ the relationship~(\ref{eq:quant}) can persist for a wide
range of parameters and so this effect suggest the use of them as new
type of frequency-to-voltage converters. A major hurdle exists,
however, in the practical application of rectification in \ssls: The
parameters for which this dynamical rectification is expected to
happen overlap exactly with the regions where the homogeneous spatial
charge distribution is unstable. In this paper, we explore a possible
way of avoiding this problem, and show that it should indeed be
possible to observe the generation of a nearly quantized \dc\ field
with realistic \ssl\ parameters.

\section{Effective circuit model with a \dc\ current source}
\begin{figure}
  \begin{center}
  \includegraphics{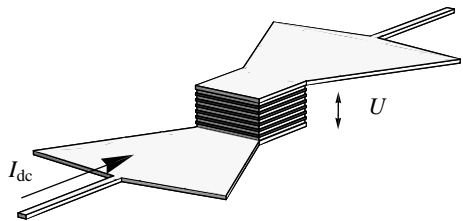}
  \end{center}
  \caption{\label{fig:schem} Schematic figure of the \ssl\ rectifier
    and the antenna. Incident radiation creates an oscillating
    potential $U_\text{ext}$ and a fixed current $\Ibias$ is supplied
    via the leads connected to the antenna.}
\end{figure}

As our model, we use an effective circuit calculation similar to
\cite{ignatov99, ghosh99, romanov00} where the active part, the \ssl,
is connected to an \ac\ circuit modeling an antenna that couples the
radiation field to the device. A schematic figure of the setup is
presented in figure \ref{fig:schem}. The external electric field
generates an oscillating contribution $U_\text{ext} = U_1 \cos \omega
t$ to the total potential across the device: $U = U_\text{ext} - \ell
E$, where $E$ is the electric field in the active region and $\ell$ is
its length. We take the amplitude $U_1$ and frequency $\omega$ of the
external potential as our given parameters. Contributions from
displacement currents are taken into account. Our new idea is to
include a circuit that acts as a direct current source, providing a
fixed current $\Ibias$. The \dc\ voltage will then be determined by a
self-consistent calculation of the model equations. We consider this
voltage as the output of the device.

The active part of the THz-rectifier is modeled using the well-known
superlattice balance equations for sinusoidal
miniband\cite{wacker02:ssreview, ignatov95}:
\begin{subequations}
  \label{eq:be0}
  \begin{eqnarray}
    \dot V &=& -\frac{e a^2 E}{\hbar^2}\left(W - \frac{\Delta}{2}\right) - \frac{1}{\tauv} V, \\
    \dot W &=& e E V - \frac{1}{\tauw}(W - \Weq). \label{eq:be-w}
  \end{eqnarray}
\end{subequations}
The variables $V$ and $W$ are the average electron velocity and
energy, respectively, where the averaging is done over a distribution
function obeying the Boltzmann transport equation. $\Weq$ is the
equilibrium energy, $\Weq = -(\Delta/2) I_1(\Delta/(2\kb T)) /
I_0(\Delta/(2\kb T))$, where $\Delta$ is the miniband width, $T$
temperature, $\kb$ Boltzmann constant, and $I_{0,1}$ are modified
Bessel functions. $E(t)$ is the total electric field that the miniband
electrons experience, and it is obtained using the equation for
current continuity across the active part of \ssl:
\begin{equation}
  \frac{\eps_0}{4\pi} \dot E + j + \frac{\ell}{A R} E = j_\text{ext}.
  \label{eq:ccont}
\end{equation}
Above, $\eps_0$ is the average \dc\ dielectric constant of the
superlattice, $j$ is the conduction electron current density across
the superlattice, $R$ is a resistance that models ohmic losses in the
superlattice, $A$ is the area of superlattice cross section, and
$j_\text{ext}$ is the external current density. This in turn is
determined by considering the effective circuit model, which gives $A
j_\text{ext} = \Ca \dot U + \Ibias$ with $\Ca$ being an effective
antenna capacitance.  The current density $j$ is given by $j = e n V$,
where $n$ is the electron density of the superlattice. Together,
\eqsref{eq:be0} and~\eqref{eq:ccont} give a closed set of first order
nonlinear ordinary differential equations.

To facilitate analysis, we scale the variables $V$ and $W$ by their
maximum values, giving $W = (\Delta/2)(1 + w)$, $V = a \Delta
(2\hbar)^{-1}\, v$.  The new variables, scaled electron velocity and
energy, $v$ and $w$ have the range $-1 \ldots +1$. Electric fields are
converted to frequency units by multiplying by $e a/\hbar$: $e a
E/\hbar = u$.  We obtain
\begin{subequations} 
  \label{eq:bes}
  \begin{eqnarray}   
    \dot v &=& -u w - \gammav v, \\
    \dot w &=& \phantom{-} u v - \gammaw (w - \weq), \\
    \dot u &=& -r^2 v - \alpha u + {\dot \phi}_\text{ext}, \label{eq:besc}  \\
    \phiext &=& \omegas \cos \omega t + i_0 t,
  \end{eqnarray}
\end{subequations}
where $r^2 = \omegapl^2/(1 + \Ca/\Cs)$, the drive amplitude $\omegas =
e a U_1/(\hbar \ell(1 + \Cs/\Ca))$, the electric field damping $\alpha
= R^{-1}(\Cs + \Ca)^{-1}$, the current bias $i_0 = a e \Ibias / (\hbar
\ell (\Cs + \Ca))$, the superlattice capacitance $\Cs = \eps_0 A /
(4\pi \ell)$ and
\begin{equation}
  \omegapl^2 = \frac{2 \pi e^2 n a^2 \Delta}{\hbar^2 \eps_0}.
\end{equation}
Here, $\omegapl$ is the plasma frequency and it describes the free
oscillation frequency of the electron gas, and also serves as the
parameter that determines the amount of nonlinearity.  Note, that
excluding the \dc\ bias current term, equations~(\ref{eq:bes}) are
formally equivalent to those used in \cite{alekseev96:dissp-chaos-ssl,
  alekseev98:spont-dc}.  For our simulations we use parameter values
that closely match those of the superlattice studied in
\cite{schomburg99}: $\Delta = 120\;\unit{meV}$, superlattice period $a
= 46\;\unit{\AA}$, number of layers $N = 130$, $n = 9\cdot
10^{17}\;\unit{cm}^{-3}$, $\gammav = 15\;\unit{THz}$, $\gammaw =
3\;\unit{THz}$, and temperature $T = 300\;\unit{K}$. In addition we
choose $\Ca = \Cs \simeq 5\;\unit{fF}$, $R = 1\;\unit{k$\Omega$}$,
roughly matching values used in \cite{romanov00}.  Denoting $\tau =
1/\sqrt{\gammav \gammaw}$, the above values give $r \tau \simeq 2.25$,
$\alpha\tau \simeq 0.015$, and $\weq \simeq -0.75$.

In this paper we then consider the \dc\ part $u_0$ of the total
electric field $u$, $u_0 = \langle u \rangle$. We would like to find
spontaneous generation of \dc\ bias, \ie\ $\langle u \rangle \neq 0$,
preferably nearly quantized so that $\langle u \rangle \simeq n
\omega$.  In the absence of the \dc\ drive current, (\ref{eq:bes})
remains invariant in the transformation $\symm$:
\begin{equation}
  (t, v, w, u) \symmto (t + T/2, -v, w, -u),
  \label{eq:symm}
\end{equation}
where $T = 2\pi / \omega$ is the period of the external \ac\ field.
If the governing equations remain invariant in transformation $\symm$,
we say that they are symmetric or have symmetry $\symm$. In our
previous works we have focused on the appearance of a spontaneous
\dc\ voltage via \emph{dynamical symmetry
  breaking}\cite{alekseev98:spont-dc, alekseev02a, alekseev05:lsslcr,
  isohatala05:sbpend, isohatala10}.  Let $u_0$ be the generated
\dc\ voltage: $u_0 = \langle u \rangle = T^{-1}\int_0^Tu(t')\;\D t'$
(we use $\langle \cdot \rangle$ to denote time averages over the
period of the function being averaged). If the equations have symmetry
$\symm$, a state with a nonzero average voltage, $u_0 \neq 0$, implies
that the solutions themselves do not follow symmetry $\symm$: $(v(t +
T/2), w(t + T/2), u(t + T/2)) \neq (-v(t), w(t), -u(t))$. This is
dynamical symmetry breaking. Here however, symmetry $\symm$ is not
present in the equations if $i_0 \neq 0$, and therefore one cannot
speak of spontaneous \dc\ following the breaking of symmetry. In spite
of this, we will show that a key feature of spontaneous generation of
a \dc\ field remains: a sharp change in the total \dc\ field near the
conditions of symmetry breaking and also the quantization of that
\dc\ component. It should be noted that the \dc\ bias current will
only be essential during the turn-on of the device. Nothing prohibits
disconnecting the bias current and restoring the symmetry of the
problem after the device has settled to a steady state.

The present work specifically addresses the problem of domain
instabilities. An underlying assumption in the models used to predict
the spontaneous \dc\ field is that the electron distribution is
spatially homogeneous. It is, however, well known that in the presence
of negative differential conductivity (\ndc), the homogeneity of the
distribution tends to be violated and domains of different electric
field strengths form\cite{ridley61, ktitorov, *ktitorov-trans,
  buttiker77}. This issue has been mainly ignored in previous
works. For stable operation of the device as a THz-rectifier we then
require that
\begin{equation}
  \frac{\partial \vdc}{\partial u_\text{dc}} > 0,
  \label{eq:stab}
\end{equation}
where $\vdc = \langle v \rangle$ is the current averaged over its
temporal period and $u_\text{dc}$ is a \dc\ probe field introduced to
the total field $u$: $u \to u + \udc$. We will refer to the steady
state values of the realized \dc\ electric field and current as the
operating point.

Condition~(\ref{eq:stab}) also effectively determines the appearance
of a spontaneous \dc\ field. Suppose $\bar v, \bar w$ is a solution to
the two first equations of~(\ref{eq:bes}) for a given field $\bar u$
containing a \dc\ probe field represented by a slow function $\udc$.
Substituting to \eqref{eq:besc} one gets ${\dot u}_\text{dc} = -r^2
\langle \bar v \rangle(u_\text{dc}) - \alpha \udc$. An equilibrium
value of $\udc$ is stable if the derivative of the right-hand side
with respect to $\udc$ is negative:
\begin{equation}
  -r^2\pd{\langle \bar v \rangle}{\udc}(\udc) - \alpha < 0.
  \label{eq:dcstab}
\end{equation}
Above, $\langle \bar v \rangle$ is the time average of $\bar v$, which
is simply the \dc\ current. Presence of the electric field damping
$\alpha$ has the effect of stabilizing the \dc\ field, however in the
$\alpha \to 0$ limit equations~(\ref{eq:dcstab}) and~(\ref{eq:stab})
coincide.  Therefore, \emph{in order to achieve spontaneous \dc\ from
  pure \ac\ excitation, the superlattice needs to be driven into
  \ndc}. In the following sections, however, we will show that a small
\dc\ bias current can make the current-voltage curve (\iv-curve) slope
positive across a range of parameters while at the same time allowing
for a substantial \dc\ voltage to appear.

As the final point regarding stability against domains, we note that
we should strengthen condition of \eqref{eq:stab}. In an experimental
setup the external \ac\ field does not rise instantly, and so during
turn-on and the initial transient phase the device might not be stable
against domains. In the absence of a model that would take into
account both formation of domains and spontaneous \dc, we resort to
comparing characteristic time scales for both processes to estimate
which will dominate. Switching times $\tau_\text{sw}$, that is, the
time it takes for a spontaneous \dc\ to appear should be faster than
the domain formation times $\tau_\text{dom}$ to avoid the interference
of the domains. In \cite{ignatov95} where a similar model of a
\ssl\ rectifier was used, $\tau_\text{sw}$ was calculated to be at
least in the range of hundreds of \ac\ drive cycles $T = 2\pi/\omega$
in the case of heavily doped wide miniband superlattices. On the other
hand, for similar superlattices the domain formation times can be as
short as fractions of
picoseconds\cite{klappenberger04:sc-domains}. Both of these
characteristic times scale down as $\omegapl$ is increased, and since
it is the dressed plasma frequency $r$, $r < \omegapl$, that here
determines the switching time, we may suppose that $\tau_\text{dom}
\lesssim \tau_\text{sw}$.  This suggests that domains are likely to
occur instead of spontaneous \dc. This problem of \ndc\ during turn-on
and transient can be solved by assuming that the external potential is
turned on slowly, and \emph{requiring that the device is stable for
  all drive amplitudes below the intended target value}. This
additional criterion bypasses all the uncertainty related to
transients and should guarantee us stable operation. Next, we will
show that in the limit $\omegapl \tau \lesssim 1$ these requirements
cannot be satisfied if a pure \ac\ drive is considered, but can be if
a small \dc\ bias current is introduced.

\section{Weak DC biased switching in the $\omegapl \tau \ll 1$ limit}
First, let us consider the case where we only include the first
harmonic of the field $u$. We assume the electric field has the form
$u = u_0 + u_1 \cos \omega t$, and neglect the dependence of $u_1$ on
our main parameters $\omegas$ and $\omega$. Also, we set $\gammav =
\gammaw = \gamma$. The first harmonic approximation is valid when the
nonlinearity is weak, that is, when $\omegapl \tau \ll 1$.

Consider the \dc\ part of the total electric field $u$. Averaging
\eqref{eq:besc} yields
\begin{equation}
  r^2 v_0(u_0, u_1, \omega) = -\alpha u_0 + i_0,
  \label{eq:udc}
\end{equation}
where $v_0(u_0, u_1, \omega)$ is the \dc\ current under \ac\ field
$u_1 \cos \omega t$. The resulting net voltage and the corresponding
current are found at the intersection of the \iv-curve and the line
$\vd$,
\begin{equation}
  \vd(u_0) = r^{-2} (-\alpha u_0 + i_0).
  \label{eq:vddef}
\end{equation}
For a given electric field $u$, $v_0(u_0, u_1, \omega)$ can be
explicitly written out as\cite{wacker02:ssreview, bass97}
\begin{eqnarray}
  v_0(u_0, u_1, \omega) = 
  \sum_{n = -\infty}^\infty
  J^2_n \left(\frac{u_1}{\omega}\right) \vet(u_0 + n\omega),
  \label{eq:iv} \\
  \vet(u) = 
  |\weq|
  \frac{\gamma u}{\gamma^2 + u^2}.
  \label{eq:vet}
\end{eqnarray}
The dependence $\vet(u)$ is the Esaki-Tsu current voltage curve
\cite{esakitsu70} and in \eqref{eq:iv} we see the familiar form of an
\ac\ irradiated superlattice where the $n$th term represents an
$n$-photon-assisted replica of $\vet$\cite{wacker02:ssreview,
  platero04:pa-xport-scns, tiengordon}.

The effect of the \dc\ bias is to shift the operating point to high
\dc\ field values as the \ac\ amplitude is increased.  This can be
seen by considering a small \dc\ voltage $u_0$.  Expanding
\eqref{eq:iv} to leading order in $u_0$ gives
\begin{equation}
  v_0(u_0, u_1, \omega) \simeq g_0(u_1, \omega)u_0,
\end{equation}
where $g_0$ is the weak \dc\ field conductivity of the \ssl\ under
\dc\ and \ac\ drives
\begin{equation}
  g_0(u_1, \omega) = 
  \sum_{n=-\infty}^\infty J^2_n\left(\frac{u_1}{\omega}\right) \vet'(n\omega)
\end{equation}
and the prime symbol means the derivative with respect to $u$.  In the
weak \dc\ field limit, stability condition~(\ref{eq:stab}) is
equivalent to $g_0(u_1, \omega) > 0$. Therefore, we wish to see how
$u_0$ changes as $g_0$ tends to zero. The generated voltage $u_0$ can
be solved from \eqref{eq:udc}:
\begin{equation}
  u_0 = \frac{i_0}{r^2 g_0(u_1, \omega) + \alpha}.
\end{equation}
Setting $g_0$ to zero one finds that $u_0 \to i_0/\alpha$, which is no
longer a small quantity even for a relatively small $i_0$, since
$\alpha \ll 1$. This suggests that before the onset of \ndc, the net
\dc\ electric field tends to large values. This is the basis of the
\dc\ biased slow switching scheme that we propose: The \dc\ bias
results in a large generated \dc\ field that appears before \ndc, thus
avoiding the domain instability where $g_0(u_1, \omega) < 0$. Clearly,
this alone is not enough, since the actual realized operating point
may in fact sit on the \ndc\ part of the \iv-curve. Whether or not
this is the case will be determined by \eqref{eq:stab}
and~\eqref{eq:udc}.

\begin{figure}
  \begin{center}
  \includegraphics{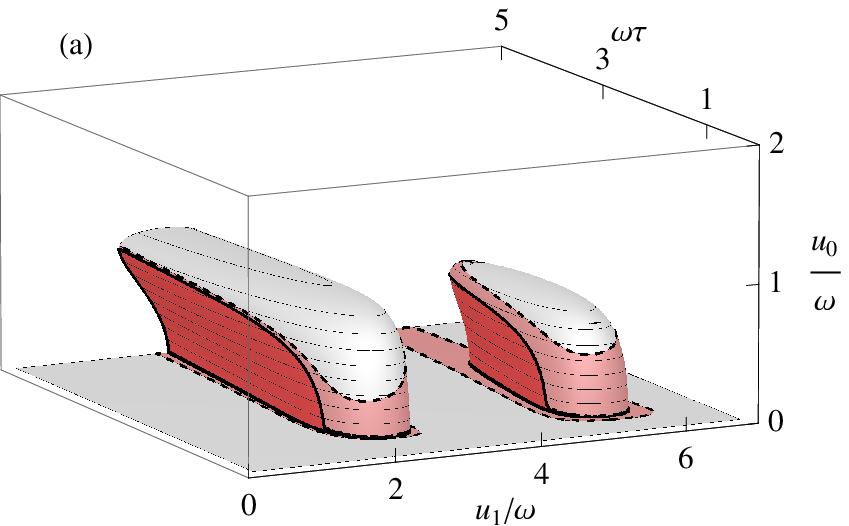}
  \includegraphics{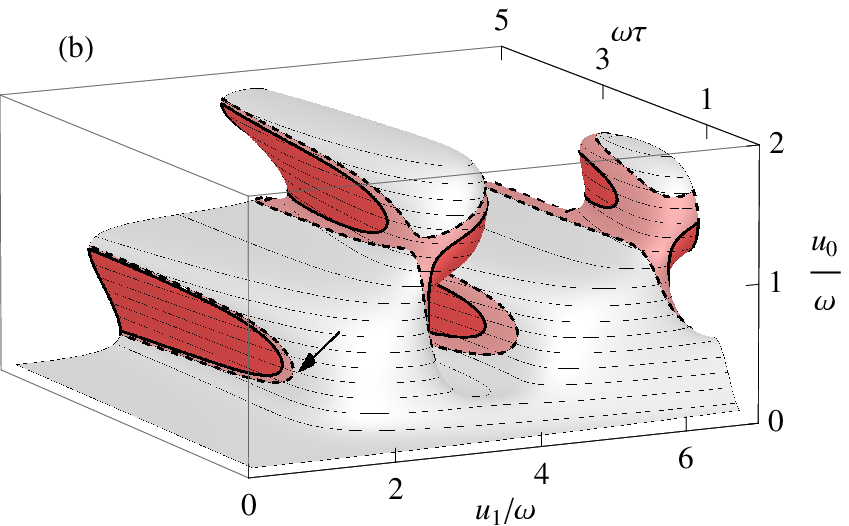}
  \end{center}
  \caption{\label{fig:3d} [Colour online] Generated \dc\ voltage as a
    function of \ac\ field amplitude $u_1$ and frequency
    $\omega$. Remaining parameters are $r \tau = 1$, $\alpha \tau =
    0.015$, and $\weq = -0.75$. Regions shaded gray show stable
    operation. Dashed lines indicate borders of \ndc\ [light red
      regions indicate \ndc\ and domains], and solid thick lines show
    points of \sn-bifurcations [dark red regions mark unstable
      branches of multistable operation]. Contours are plotted at
    level values of $u_0/\omega$ and are spaced at intervals of
    $0.1$. (a) Bias current off, $i_0 = 0$. Two regions of spontaneous
    \dc\ are visible, however, they are flanked by regions unstable
    operation by all sides. (b) Bias current on, $i_0 = \iopt$.
    Strong response in $u_0$ is seen where spontaneous \dc\ appeared
    in the unbiased case. Arrow at $u_1/\omega\approx 1.70$, $\omega
    \tau \approx 1.62$, $u_0/\omega \approx 0.39$ marks the point
    where \ndc\ first appears. Stable operation with large $u_0$ is
    then possible for $\omega \tau \lesssim 1.6$.}
\end{figure}

Let us now consider the frequency dependence of $u_0$. In the high
frequency $\omega \tau \gg 1$ limit the \dc\ voltage can be very
nearly quantized when the bias current is absent\cite{ignatov95,
  romanov00}. This does not change when the \dc\ drive is
included. Since near $u_0 = n \omega$ the $n$ photon term dominates
the series expression~(\ref{eq:iv}), the current is approximately
given by $J^2_n(u_1/\omega)\vet(u_0 - n \omega)$. Using \eqref{eq:udc}
and assuming $|u_0 - n \omega| \ll 1$ we find that
\begin{equation}
  u_0 \simeq n \omega 
  + \gamma \frac{i_0 - \alpha n \omega}
                {\alpha \gamma + |\weq| J_n(u_1/\omega)^2r^2}.
                \label{eq:u0hf}
\end{equation}
The correction term is small since we take $i_0\tau^2, \alpha\tau \ll
1$, and so the realized operating point voltage is nearly quantized.
Further, this point is always stable provided $J_n(u_1/\omega)$ is not
too close to zero. Note also, that \eqref{eq:u0hf} gives an
approximate optimal drive current for observing quantized $u_0$.  The
correction term vanishes and the quantization is improved by choosing
$i_0 = \iopt$ where
\begin{equation}
  \label{eq:iopt}
  i_\text{opt} = n \omega \alpha.
\end{equation}
The high-frequency limit is also particularly susceptible to domain
instabilities during turn-on. This is because of hysteretic response
to changing drive amplitude.  The current replicas, that is, the
individual terms in \eqref{eq:iv}, define local peaks in the
\iv-curve. Since the slope of $\vd$ is low, the intersections between
$\vd$ and $v_0$ tend to come in pairs, one on each side of a local
maximum of current.  These intersections define a pair of operating
points, one stable and the other unstable. Consider one such pair near
$u_0 = k \omega$. Should $J_k(u_1/\omega)$ tend to zero as $u_1$
increases, that local current peak dips below $\vd$ and the operating
points collide and are subsequently destroyed. This is essentially a
saddle-node (\sn) bifurcation and causes a problem since
\sn\ bifurcations always occur in conjunction with \ndc. One can see
this easily by noting that \sn-bifurcation implies a double root of
\eqref{eq:udc}, which in turn means that slopes of $v_0$ and $\vd$
must there agree:
\begin{equation}
  \pd{\vdc}{\udc} = - \frac{\alpha}{r^2}.
  \label{eq:sn}
\end{equation}
Clearly then, at an \sn-bifurcation the system is already in \ndc.
Since equilibria are destroyed and created at \sn-bifurcations, their
presence implies multistability and hysteretic response.  Requiring
stability against formation of domains therefore means that the
voltage $u_0$ response to changing $u_1$ must be non-hysteretic. This
effectively sets an upper limit to the frequencies by which stable
turn-on can be guaranteed.

In the low-frequency limit, $\omega\tau\ll 1$, the current
adiabatically follows the unirradiated \iv-curve \eqref{eq:vet}. The
\dc\ current is then approximately given by $\langle \vet(u_0 + u_1
\cos \omega t) \rangle$. In this case, $u_0$ as given by
\eqref{eq:udc} is always nearly zero, and exactly zero in the absence
of the bias current. The low and high frequency limits together
demonstrate that there is a band of frequencies near $\omega \tau \sim
1$ whereby strong response of $u_0$ can be observed without
domains. In accordance with \eqref{eq:u0hf}, the value of $u_0$ should
be closest to being quantized in the high end of that range of
frequencies.

To see what effect the \dc\ bias current has on the parameters where
rectification and quantized electric fields appear, we have in
figure~\ref{fig:3d} plotted the \dc\ voltage obtained by solving
\eqref{eq:udc} as a function of $u_1$ and $\omega$.  We have chosen
the parameter values as $r \tau = 1, \weq = -0.75, \alpha \tau =
0.015$. Borders of regions where \ndc\ appears are indicated by dashed
lines. Thick black curves mark \sn-bifurcations. In figure
\ref{fig:3d}(a) the bias current is off. Two regions where spontaneous
\dc, $|u_0| > 0$, can be seen near $u_1/\omega \simeq 2.4$ and
$u_1/\omega \simeq 5.5$, the two first roots of Bessel $J_0$. The
spontaneous \dc\ forms plateaus satisfying the near-quantization
condition~(\ref{eq:quant}) in the high frequency end of the
figure. The problem of domain instabilities is also evident. The
regions of spontaneous \dc\ are surrounded by regions of \ndc. For low
but non-zero values of $u_0$ \ndc\ is still present, but disappears
for $u_0$ closer to $\omega$. Although the nearly quantized values of
$u_0$ are stable, they can only be reached via a hard-mode excitation,
\ie\ by instant turn-on and an initial electric field that is already
close to the quantized value. With the more realistic adiabatic
turn-on scheme, spontaneous \dc\ appears impossible.

Turning on the \dc\ bias current essentially different behaviour is
seen. This is shown in figure~\ref{fig:3d}(b) where \dc\ bias current
is set to $i_0 = \iopt$ (optimizing for the $u_0 = \omega$ plateau,
\ie\ setting $n = 1$ in \eqref{eq:iopt}).  The plateaus of nearly
quantized spontaneous \dc\ are still present where $\omegas/\omega$ is
close to the first two Bessel $J_0$ roots. Additional plateaus closer
to $u_0 = 2 \omega$ have now also appeared. The \dc\ electric field is
non-zero for other values of the parameters as well, but nonetheless
there is a sharp increase or decrease in $u_0$ where spontaneous
\dc\ appeared or disappeared in the pure \ac\ case above. Importantly,
this time \ndc\ is suppressed and stable operation is possible for a
range of $\omegas$ from $0$ to $2.4$, where the dependence $u_0 \sim
\omega$ can be seen.  \ndc\ first appears at frequency $\omega \tau
\approx 1.62$ (indicated by an arrow in the figure, $u_1 \approx 1.70
\omega$) followed by multistability at $\omega \tau \approx 1.80$,
$u_1 \approx 1.68 \omega$.  This region extends long into higher
frequencies and so it impedes reaching the quantized voltage
plateau. Restricting to $\omega \tau \lesssim 1.6$ the quantization is
not so good, since well-developed plateaus are not present. Avoiding
domains does then clearly limit how good quantization can be reached.

\section{Numerical simulation, $\omegapl \tau \gtrsim 1$}
We now turn to the case of $\omegapl\tau \gtrsim 1$. Now all harmonics
of the electric field are included and our results are parametrized by
the applied \ac\ potential amplitude $\omegas$ and frequency
$\omega$. The problem is studied by numerically solving \eqref{eq:bes}
for varying $\omega$ and $i_0$.  We would like to find conditions for
strong response in the \dc\ voltage without \ndc\ or hysteresis, and
further, preferably quantized values of $u_0$. These requirements
translate to having $u_0$ dependence on $\omegas$ resembling a step
function. Results of the $\omegapl \tau \lesssim 1$ limit showed that
quantization becomes better as frequency is increased, but that there
is an upper limit in $\omega \tau$ that still yields stable operation.
We will use that finding as a guide and aim to find the highest
applied field frequencies that are free from \ndc\ and hysteresis. The
$\omega \tau$ upper limits turn out to be higher in the $\omegapl \tau
\gtrsim 1$ case.

\begin{figure}
  \begin{center}
  \includegraphics{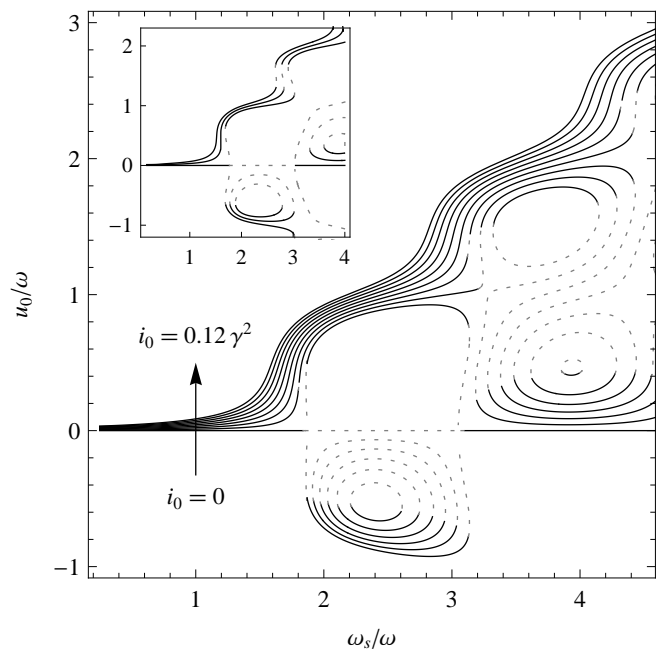}
  \end{center}
  \caption{\label{fig:3} Generated \dc\ voltage as a function of the
    drive amplitude at different bias currents. Solid curves indicate
    stable values while dashed curves operating points with
    \ndc. Inset axis are the same as in main figure. Main figure:
    Parameters correspond to existing highly doped, wide miniband
    \ssls: $\omega \tau = 2.0$, $r \tau = 2.25$, $\gammav\tau = 2.0$,
    $\gammaw \tau = 0.5$, $\alpha \tau = 0.015$, and $\weq = -0.75$. A
    range of bias currents is from $i_0 = 0$ to $i_0 = 0.12
    \gamma^2$. Inset: Parameters are pushed to higher but reasonable
    values, $\omega \tau = 3.0$, $r \tau = 5.0$, $\gammav\tau = 2.0$,
    $\gammaw \tau = 0.5$, $\alpha \tau = 0.005$, and $\weq =
    -0.75$. Bias currents range from $i_0 = 0$ to $0.39 \gamma^2$.}
\end{figure}

We will start by considering a highly doped, wide miniband
superlattice that is currently realizable. In figure~\ref{fig:3}
sample runs are presented where nearly quantized \dc\ voltage can be
achieved for parameters corresponding to such an \ssl\ device. In the
main figure, the values were chosen as $r \tau = 2.25$, $\omega \tau =
2.0$, $\gammav \tau = 2.0$, $\gammaw \tau = 0.5$, $\alpha \tau =
0.015$, and $\weq = -0.75$.  A range of \dc\ currents $i_0$ from $0$
to $0.12 \gamma^2$ was used to demonstrate the effect of current
biasing. The $i_0 = 0$ curve shows again a region of spontaneous
\dc\ near $\omegas / \omega \simeq 2.4$. This is surrounded as
expected by \ndc\ and here further by multistability and so stable
operation is not available. As in the $\omegapl \tau \lesssim 1$ case,
the inclusion of the \dc\ bias removes the \ndc\ and the
multistability. For small but non-zero $i_0$ the \ndc\ is not
completely gone, but does disappear quickly as the bias is increased.
A rapid rise in $u_0$ is observed near the symmetry breaking
bifurcation of the $i_0 = 0$ curve ($\omegas / \omega \simeq 1.8$),
and further, this sudden increase quickly shoulders off to form a
slightly tilted plateau where $u_0 \simeq \omega$.  The realized $u_0$
depends only weakly on $i_0$, showing that the device is not sensitive
to the choice of bias current as long as it is sufficiently high to
remove the instabilities. Higher near-quantized plateaus for which
$u_0/\omega \simeq n$ with $n > 1$ also become available, though now
the quantization is less good.

The inset in figure~\ref{fig:3} shows what can be achieved with a
reasonable push in the parameters. Here, we have boosted the plasma
frequency to $r \tau = 5$ and drive frequency to $\omega \tau = 3.0$
while reducing the inverse parallel resistance $\alpha$ to $0.005
\gamma$. Dependence on the bias current is again similar (here, $i_0 =
0 \ldots 0.39 \gamma^2$) with the case of zero external current
showing instabilities that are removed as $i_0$ is increased. With
these parameters the shoulder where $u_0$ levels off to $\simeq
\omega$ is sharper and the quantization better, yet $u_0$ is single
valued in $\omegas$ as desired. Quantization with $\eta \approx 0.95$
($u_0 = \eta \omega$) can be observed here for a wide range of
$\omegas$.

Note that the $\omegapl \tau \lesssim 1$ limit suggests a scaling
law. The equations that determine $u_0$, \eqref{eq:udc}
and~\eqref{eq:iv}, depend on $r$, $\alpha$, and $i_0$ only via the
fractions $\alpha/r^2$ and $i_0/r^2$. Therefore, one could hypothesize
that the increased stability of $\omegapl \tau \gtrsim 1$ case is only
due to a decrease in the effective inverse parallel resistance
$\alpha/r^2$, and could then be replicated by keeping $r$ fixed and
decreasing $\alpha$ instead. We tested this hypothesis by running
simulations with fixed $\alpha/r^2$ and varying $r$ and found it to be
false, so that increasing $r$ really does have an effect of
suppressing \ndc.

We see that stable operation at higher applied field frequencies can
be obtained in the limit of a high plasma frequency, or a high doping
and a wide miniband. Because higher $\omega \tau$ can be reached, also
the quantization of the spontaneous \dc\ voltage is
improved. Nonetheless, \ndc\ and hysteretic response still set strict
limits to frequency and hence to the quantization that can be
achieved. Stability against domains must be balanced against how small
$|u_0 - n \omega|$ is desired. To elaborate further on the parameters
that affect the spontaneous \dc, we note that there is some
sensitivity to the value $\alpha$. Zero inverse parallel resistance,
$\alpha = 0$, results in a more hysteretic response, and so a small
degree of damping in the equation for $u$ is preferable. However, for
$\alpha \tau \gtrsim 0.1$ the generated \dc\ voltage becomes close to
zero, so smallness of $\alpha$ is still required. In terms of the
\dc\ bias current, the device does not appear to be very sensitive to
its value as long as it is sufficient to remove \ndc\ near zero
\dc\ voltages.  Increasing $i_0$ further to moderate values has little
effect on stability, it only shifts the point where strong response
occurs to lower values of $\omegas / \omega$.  Again, we note that
since the spontaneous \dc\ states are often stable when $i_0 = 0$, the
bias current is not necessary anymore once a stable equilibrium is
reached provided of course that the unbiased operating point is not in
\ndc.

\section{Conclusions}
We have presented a turn-on scheme for a semiconductor superlattice
THz-rectifier that addresses the problem of stability against
formation of electric domains due to \ndc.  While a spontaneous
\dc\ voltage state of a pure \ac\ driven semiconductor superlattice is
stable, this voltage state can only be reached by driving the device
into conditions where domains do form. Our proposed solution to this
problem is to apply a weak \dc\ current bias while the applied field
is adiabatically increased to its target value. The external \dc\ bias
has the effect of shifting the operating point towards higher
\dc\ voltages whereby the device shows positive differential
conductivity, and therefore is stable against formation of domains.
This shifting appears as a strong response in the \dc\ voltage as a
function of the amplitude of the external \ac\ field and is seen for
parameters that correspond to dynamical symmetry breaking in the pure
\ac\ driven case.

For typical superlattices and experimentally realizable parameters of
applied field, together with small \dc\ bias currents, the strong
response in the \dc\ voltage and stable switching to a spontaneous
\dc\ field state are indeed found. Appearance of \ndc\ and hysteretic
response however do still set an upper limit to the applied radiation
(angular) frequencies $\omega$ that give the desired behaviour. For
weak nonlinearity, $\omega$ is roughly limited by $1.5 \sim 2$ times
the characteristic scattering rates. Somewhat counter-intuitively, in
the limit of strong nonlinearity (high plasma frequencies) stable
operation can be achieved for higher drive frequencies. In accordance
with previous results on generation of spontaneous \dc\ voltage, we
found the generated \dc\ field becomes closer to an integer multiple
of $\hbar \omega / e a$ as $\omega$ is increased. Therefore, our
results show that for optimal, stable frequency to voltage conversion,
high plasma frequencies are preferable. Our numerical results show
that good quantization of the generated voltage can be achieved for
realistic semiconductor superlattices. These findings demonstrate
that it may indeed be possible to apply semiconductor superlattices as
room temperature frequency-to-voltage converters and detectors of a
strong THz radiation.

\ack
We are thankful to Erkki Thuneberg for a critical reading of the
manuscript and valuable remarks concerning the effective circuit
description.

\section*{References}
\mciteSetMidEndSepPunct{\par\hspace{\itemindent}}{}{\relax}
\bibliographystyle{iopartmbib}
\bibliography{bibliography}

\end{document}